\documentstyle[aps,pre]{revtex}

\makeatletter
\def\appendix{\par\clearpage
  \setcounter{section}{0}
  \setcounter{subsection}{0}
  \@addtoreset{equation}{section}
  \def\@sectname{Appendix~}
  \def\theequation{\thesect.\arabic{equation}}
  \def\thesect{\Alph{section}}
  \def\thesection{\@sectname\Alph{section}}}
\makeatother

\newcommand{\beq}{\begin{equation}}
\newcommand{\eeq}[1]{\label{#1} \end{equation}}
\newcommand{\beqar}{\begin{eqnarray}}
\newcommand{\eeqar}[1]{\label{#1} \end{eqnarray}}
\newcommand{\insertplot}[1]{
\centerline{\psfig{figure={#1},height=11.0cm}}
}
\newcommand{\insertplotshot}[1]{
\centerline{\psfig{figure={#1},height=9.0cm}}
}
\newcommand{\dlt}{\bigtriangleup}
\begin{document}
\input{psfig.sty}

\begin{titlepage}
\vskip 0.3 cm
\centerline{\bf THE INITIAL STATE OF
ULTRA-RELATIVISTIC HEAVY ION COLLISION} 
\vskip 1 cm

\centerline{V.K. Magas$^{1,\star}$, L.P. Csernai$^{1,2,\star}$ 
and D.D. Strottman $^{3,4,\diamond}$} 
\vskip 0.5 cm

\centerline{$^{1}$ \sl Section for Theoretical and Computational Physics,
 Department of Physics} 
\centerline{\sl University of Bergen, Allegaten 55, N-5007, Norway}
\vskip .3cm
\centerline{$^{2}$ \sl KFKI Research Institute for Particle and Nuclear
Physics} 
\centerline{\sl P.O.Box 49, 1525 Budapest, Hungary}
\vskip .3cm
\centerline{$^{3}$ \sl Theoretical Division, Los Alamos National Laboratory} 
\centerline{\sl Los Alamos, NM, 87454, USA}
\vskip .3cm
\centerline{$^{4}$ \sl Institut f\"ur Theoretische Physik, 
Universit\"at Frankfurt} 
\centerline{\sl Robert-Mayer-Str. 8-10, D-60054 Frankfurt am Main, Germany}

\vskip 0.5cm

\begin{abstract}
A model for energy, 
pressure and flow velocity distributions at the beginning of
ultra-relativistic heavy ion collisions is presented, which can be used as
an initial condition for hydrodynamic calculations.  Our model takes into
account baryon recoil for both target and projectile, arising from the
acceleration of partons in an effective field, $F^{\mu \nu}$, produced in
the collision. The typical field strength (string tension) for RHIC
energies is about $5-12 \ GeV/fm$, what allows us to talk about "string
ropes".  The results show that a QGP forms a tilted disk, such that the
direction of the largest pressure gradient stays in the reaction plane,
but deviates from both the beam and the usual transverse flow directions.
Such initial conditions may lead to the creation of "antiflow" or "third
flow component" \cite{CR}. \linebreak
\end{abstract}

\vskip .5cm
\hrule
\noindent

$\begin{array}{ll}
^{\star}\mbox{{\it email address:}} &
   \mbox{VLADIMIR,CSERNAI~@FI.UIB.NO}
\end{array}
$

$\begin{array}{ll}
^{\diamond}\mbox{{\it email address:}} &
   \mbox{DDS~@LANL.GOV}
\end{array}
$

\vfill
\end{titlepage}
\eject
\textheight 210mm
\topmargin 2mm
\baselineskip=24pt

\section{Introduction}

\label{one}

Fluid dynamical models are widely used to describe ultra-relativistic
heavy ion collisions.  Their advantage is that one can vary flexibly the
Equation of State (EoS) of the matter and test its consequences on the
reaction dynamics and the outcome. For example, the only models which may 
handle the supercooled QGP are hydrodynamical models with corresponding EoS. 
In energetic collisions of large heavy
ions, especially if a Quark-Gluon Plasma (QGP) is formed in the collision,
one-fluid dynamics is a valid and good description for the intermediate
stages of the reaction. Here, interactions are strong and frequent, so
that other models, (e.g. transport models, string models, etc., that assume
binary collisions, with free propagation of constituents between
collisions) have limited validity. On the other hand, the initial and
final, Freeze-Out (FO), stages of the reaction are outside the domain of
applicability of the fluid dynamical model.

Thus, the realistic, and detailed description of an energetic
heavy ion reaction requires a Multi Module Model, where the different
stages of the reaction are each described with a suitable theoretical
approach. It is important that these Modules are coupled to each other
correctly: on the interface, which is a three dimensional hyper-surface in
space-time with normal $d\sigma^\mu$, all conservation laws should be
satisfied, e.g. $[T^{\mu\nu}d\sigma_\nu] = 0$ (here 
the square brackets means difference between 
new and old phases or modules), and entropy should not
decrease, $[S^\mu d\sigma_\mu] \ge 0$. These matching conditions were
worked out and studied for the matching at FO in detail in Refs.
\cite{FO1,FO2,FO2a,FO3,FO3a,FO3b}.

We would like to discuss the  entropy condition 
in more detail. Obviously, the number of degrees of freedom and 
correspondingly the entropy density is  reduced during 
hadronization process. So, how can we  avoid decreasing of the entropy? 
Two scenarios have been proposed. The first one is the gradual hadronization 
scenario i.e., the hadronization is so slow that during this process the volume 
of the system becomes considerably larger to compensate for the reduction of 
entropy density. If this would be so, our long living, gradually expanding
 QGP should 
be observed in HBT experiments, e.g., as a peak in the $R_{out}/R_{side}$
ratio \cite{RG96}. The preliminary data from STAR and PHENIX
do not support this scenario \cite{LJP_qm}. The second possibility is
the fast hadronization from supercooled QGP \cite{CM95}. This hypothesis 
can be checked 
only in hydrodynamical models, which use the EoS as  direct input.

After hadronization and FO, matter is already dilute and  
can be described well with kinetic models.

The initial stages are more problematic. Frequently, two or three fluid
models are used to remedy the difficulties and to model the process of QGP
formation and thermalization \cite{A78,C82,bsd00}. Here the problem is
transferred to the determination of drag-, friction- and transfer- terms
among the fluid components, and a new problem is introduced with the
(unjustified) use of an EoS in each component in a nonequilibrated
situations where an EoS is not defined. Strictly speaking this approach can
only be justified for mixtures of noninteracting ideal gas components.
Similarly, the use of transport theoretical approaches assuming dilute
gases with binary interactions is questionable, as, due to the extreme
Lorentz contraction in the Center of Mass (CM) frame, enormous particle 
and energy
densities with the immediate formation of a perturbative vacuum should be
handled. Even in most parton cascade models these initial stages of the
dynamics are just assumed in form of some initial condition, with little
justification.

Our goal in the present work is to construct a model, based on the recent
experiences gained in string Monte Carlo models and in parton cascades.
One important conclusion of heavy ion research in the last decade is that
standard 'hadronic' string models fail to describe heavy ion experiments.

All string models had to introduce new, energetic objects: string ropes
\cite{bnk84,S95}, quark clusters \cite{WA96}, or fused strings \cite{ABP93},
in order to describe the abundant formation of massive particles like
strange antibaryons.  Based on this, we describe the initial moments of
the reaction in the framework of classical (or coherent) Yang-Mills
theory, following Ref. \cite{GC86} assuming a larger field strength (string
tension) than in ordinary hadron-hadron collisions.  For example,
calculations both in the Quark Gluon String Model (QGSM) 
\cite{ASC91pl,ASC91prl,ACS92}
and in the Monte Carlo string fusion model \cite{ABP93} 
indicate that the
energy density of strings reaches $8\ -\ 10\ GeV/fm$ already in SPS
reactions, nearly $10$ times more than 
the tension used in 
standard, 'hadronic', string models where $\sigma \approx 1\ GeV/fm$.  
In addition we now satisfy all conservation laws exactly,
while in Ref.  \cite{GC86} infinite
projectile energy was assumed, and so, overall energy and momentum
conservation was irrelevant.  Thus, in this approach for the first time
the initial transparency/stopping and energy deposited into strings and
"string ropes" will be determined consistently with each other.  
Recent parton kinetic models \cite{EK99,DG00} indicate that 
quark and gluon density saturation takes place in a very short time - 
$\tau_{sat}=0.09-0.27\ fm/c$ for LHC - SPS energies \cite{DG00}, 
and pressure needs $\tau_p=1-5\ fm/c$ to be established \cite{EK99}.
More importantly the first experiments at RHIC yield strong elliptic flow,
which cannot be reproduced in any other model, except in fluid dynamical models
with QGP EoS \cite{QM01}. This is a strong experimental 
indication that transverse pressure builds up early in these reactions, 
in a few $fm/c$, and strong stopping is also necessary 
to create strong flow before Freeze Out, which usually happens when 
the system size is not more than $10\ fm$. 	
We present initial conditions for $t_{lab}=2-5\ fm/c$, which is in 
agreement with previous estimations as well as with data.

We do not
solve simultaneously the kinetic problem leading to parton equilibration,
but assume that the arising friction is such that the heavy ion system
will be an overdamped oscillator, i.e.,  yo-yoing of the two heavy ions
will not occur, as all recent string and parton cascade results indicate.

\section{Formulation of model}
\label{two}

Our basic idea is to generalize the model developed in \cite{GC86} for
collisions of two heavy ions and improve it by strictly satisfying
conservation laws \cite{MCS00,CAM00,MCS00-1}. 
First of all, we would create a grid
in the $[x,y]$ plane ($z$ is the beam axes, $[z,x]$ is the reaction
plane).  We will describe the nucleus-nucleus collision in terms of
steak-by-streak collisions, corresponding to the same transverse
coordinates, $\{x_i, y_j\}$.  We assume that baryon recoil for both target
and projectile arise from the acceleration of partons in an effective
field $F^{\mu\nu}$, produced in the interaction.  Of course, the physical
picture behind this model should be based on chromoelectric flux tube or
string models, but for our purpose we consider $F^{\mu\nu}$ as an
effective Abelian field. 
The most important consequence of the non-Abelian fields, i.e., its
self interaction and the resulting flux tubes of constant cross section,
are, nevertheless, reflected in our model:
assuming that the field is one-dimensional. The fields generated 
by the colliding streaks are of constant cross section during the whole
evolution, and only their lengths increase with time. As the string tension 
is constant, the energy of the string increases linearly with its increasing 
length.  
The single phenomenological parameter we use to
describe this field must be fixed from comparison with experimental data.

We describe the streak-streak collision using conservation laws:
\beq
\partial_\mu \sum_i T_i^{\mu\nu}=\sum_i F_i^{\nu\mu} q_i N_{i \mu} \ ,
\eeq{eq1}
\beq
\partial_\mu \sum_i N_i^\mu = 0 \ , \quad i=1,2\  ,
\eeq{eq2}
where $N_i^\mu$ is the baryon current of $i$th nucleus, $q_i$ is the color
charge and it will be discussed in more detail later.  We are working in
the Center of Rapidity Frame (CRF), which is the same for all streaks.
The concept of using target and projectile reference frames has no
advantage any more.  We will use the parameterization:
\beq
N_i^\mu=n_i u_i^\mu \ ,
\quad
u_i^\mu=(\cosh y_i,\ \sinh y_i)  \ .
\eeq{eq3}
$T^{\mu\nu}$ is the energy-momentum flux tensor. It
consists of five parts, corresponding to both
nuclei and free field energy (also divided into two parts) and one 
term defining the QGP perturbative vacuum.
\beq
T^{\mu\nu}=\sum_i T_i^{\mu\nu}+T^{\mu\nu}_{pert}=
\sum_i\left[ e_i\left(\left(1+c_0^2\right)u_i^\mu u_i^\nu
- c_0^2g^{\mu\nu}\right)
+T_{F,i}^{\mu\nu}\right]+B g^{\mu\nu}\ ,
\quad i=1,2 \ .
\eeq{eq5}
Here $B$ is the bag constant, the equation of state is $P_i=c_0^2 e_i$, 
where $e_i$ and $P_i$ 
are energy density and pressure of QGP.

Within each streak we form only one flux tube with a uniform field strength
or field tension, $\sigma$, from the target to the projectile. For practical
purposes we, however, divide this field into two spatial domains, a
target and a projectile domain, ($i= 1,2$), separated at a fixed point,
$z_{sep}$, so that $\sigma_1 = \sigma_2 = \sigma$. 
The choice if this point will be specified later. (The field
is constant and the only change is that it extends with time at its two
ends.)

In complete analogy to electro-magnetic field
\beq
F_i^{\mu\nu}=\partial^\mu A_i^{\nu}-\partial^\nu A_i^{\mu}=\left(
\begin{array}{cc}
0 & -\sigma_i \\
\sigma_i & 0
\end{array}\right) \ \ ,
\eeq{eq6}
\beq
\sigma_i=\partial^3 A_i^{0}-\partial^0 A_i^{3}\ ,
\eeq{eq7}
\beqar
T_{F,i \mu\nu}=-g_{\mu\nu}{\cal L}_{F,i}+\sum_\beta \frac{{\cal L}_{F,i}}
{\partial \left(\partial^\mu A_i^{\beta}\right)}\partial_\nu A_i^{\beta}
\ \ ,
\eeqar{eq8}
\beq
{\cal L}_{F,i} = - \frac{1}{4}F_{i \mu\nu}F_i^{\mu\nu}\ .
\eeq{eq9}

In our case the string tensions, $\sigma_i$, will be
constant in the space-time region after string creation and before string
decay. The creation of fields will be discussed later in more detail. 

To get the analytic solutions of the above equations, we
use lightcone variables
\beq
(z,t)\rightarrow(x^+,x^-),\quad x^\pm=t\pm z \ .
\eeq{eq12}
Following \cite{GC86}, we insist that $e_1, y_1,  n_1$ 
are functions of $x^-$ only and
 $e_2, y_2,  n_2$ depend on $x^+$ only.

In terms of lightcone variables:
\beq
N_i^\pm=N_{i, \mp}= n_i(u_i^0\pm u_i^3)= n_i e^{\pm y_i} \ ,
\eeq{eq13}
\beq
T_i=\left(
\begin{array}{cc}
T_i^{++} & T_i^{+-} \\
T_i^{-+} & T_i^{--}
\end{array}\right)=\frac{1}{2} 
\left(
\begin{array}{cc}
h_{i+} e^{2 y_i} & h_{i-} \\
h_{i-} & h_{i+} e^{-2 y_i}
\end{array}\right) + T_{F,i}\ ,
\eeq{eq14}
where
\beq
h_{i+}=(1+c_0^2)e_i\ ,
\quad
h_{i-}=(1-c_0^2)e_i\ .
\eeq{eq16}
The other tensors in the light cone variables are:
\beq
F_i=\left(
\begin{array}{cc}
F_i^{++} & F_i^{+-} \\
F_i^{-+} & F_i^{--}
\end{array}\right)= 
\left(
\begin{array}{cc}
0 & \sigma_i \\
-\sigma_i & 0
\end{array}\right) \ .
\eeq{eq17}
\beq
T_{pert}=\left(
\begin{array}{cc}
0 &  B \\
B & 0
\end{array}\right) \ .
\eeq{eq17a}
The energy-momentum tensor for free field in the light cone variables is: 
\beq
T_{F,i}=\frac{1}{2}
\left(
\begin{array}{cc}
\sigma_i^2  & 0 \\
0 & \sigma_i^2 
\end{array}\right) \ .
\eeq{eq22}

\begin{figure}[htb]
\centerline{\psfig{figure={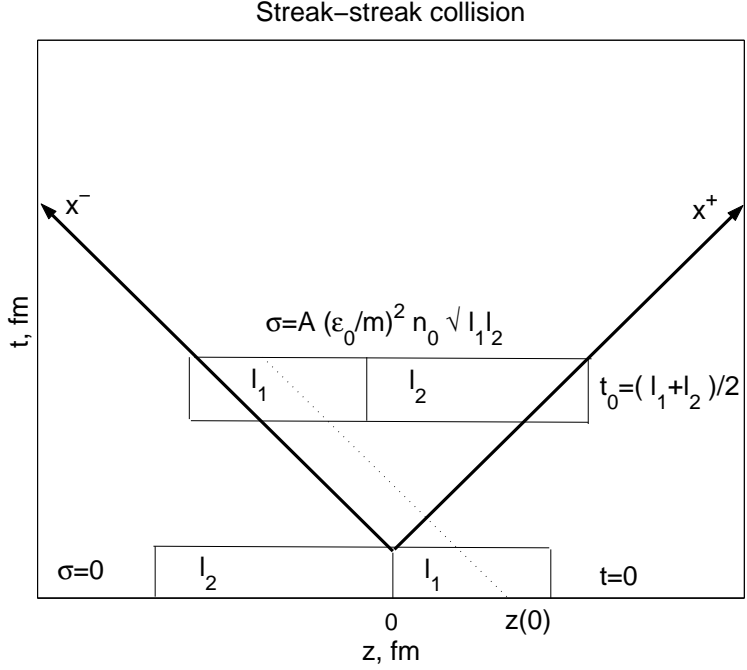},height=9.0cm}}
\vspace{0.6cm}
\caption[]{Streak-streak collision. $t=0$ at the time of first 
touch of streaks.
$t=t_0$ corresponds to complete penetration of streaks
through each other. At this time strings are completely created, i.e.,
string tension reaches an
absolute value $\sigma=A\left(\frac{\varepsilon_0}{m}\right)^2 n_0
\sqrt{l_1l_2}$ (\ref{eq44}).}

\label{fig1}
\end{figure}

At the time of the first touch of two streaks, $t=0$, there is no string
tension.  We assume that strings are created, i.e., the sting tension
achieves the value $\sigma$ at time $t=t_0$, corresponding to complete
penetration of streaks through each other (see Fig. \ref{fig1}).

\section{Conservation laws and string creation}

\label{three}

In lightcone variables eq. (\ref{eq2}) may be rewritten as
\beq
\partial_-N_1^-+\partial_+N_2^+=0 \ .
\eeq{eq19}
So, we have a sum of two terms, each depending on different 
independent variables,
and the solution can be found in the following way:
\beq
\begin{array}{ll}
\partial_-N_1^-=a, & \partial_+N_2^+=-a \ , \\
N_1^-=a x^- + (N_1^-)_0, & N_2^+=-a x^+ + (N_2^+)_0 \ ,
\end{array}
\eeq{eq20}
where the index $)_0$ indicates the initial proper density, which is 
the normal nuclear density:  $n_0=0.145\ fm^{-3}$.
Since both $N_1^-$ and $N_2^+$ are positive 
(and also more or less symmetric) 
we can conclude that for
our case $a=0$.

Finally
\beq
N_1^-= n_1 e^{-y_1}= n_0 e^{y_0} \ , \quad
N_2^+= n_2 e^{y_2}= n_0 e^{y_0} \ ,
\eeq{eq21}
\beq
 n_1= n_0 e^{y_0+y_1}\ , \quad   n_2= n_0 e^{y_0-y_2}  \ .
\eeq{eq21a}
where $y_0\ (-y_0)$ is the initial rapidity of nucleus $2\ (1)$ in the center 
of rapidity frame (CRF), respectively. The other components are given by 
eq. (\ref{eq13}).

Let us make analogy to electro-magnetic field, where two charges $q_1$ 
and $-q_2$ move in the opposite directions, creating string-like field between 
them, $\vec{E}=(0,0,E)$, which is constrained transversally into a 
constant cross section.
 The Z-axis goes through charges $q_1$ 
and $-q_2$ and directed from  $q_1$ 
to $-q_2$ (let us assume that we have such a field configuration).
So, forces acting on our charges, $q_1 E$ and $-q_2 E$, have opposite sign 
and both are working against the expansion of the "string". 
In our effective model we use color charges,
and assume that the vectors of these color charges point 
in the opposite directions in the color space \cite{ABP93}, so that
the forces acting on both target and projectile partons are opposite 
both stopping 
the expansion of the streak. 
As our field strength (string tension, $\sigma$) is not yet defined
we normalize the charges to unity:
\beq
q_1=-q_2=1>0\ , \quad {\rm while} \quad \sigma_1=\sigma_2=\sigma\ . 
\eeq{sch}
Then we have the forces acting in $z$ direction: $q_1\sigma_1=\sigma\ , 
 $  and $\ q_2\sigma_2=-\sigma$. 
Notice again that after string creation fields $\sigma_1(x)$ and $\sigma_2(x)$ 
are spatially separated as well as the baryon densities, 
$n_1$ and $n_2$; i.e., after complete penetration of 
the initial streaks through each other (see Fig. 1), $\sigma_2$ acts on the 
partons on the right side of the separating point $z_{sep}=(l_1-l_2)/2$,     
while the $\sigma_1$ acts on those on the left side. In the absence of matter, 
in the middle both fields are identical, so the exact position of the 
separating point does not play any role until it does not enter 
the target or projectile matter.
The fields $\sigma_1$ 
and $\sigma_2$ are generated by the corresponding 4-potentials $A_i$, 
which are different and spatially separated in the same way. 

As it was described above we do not generate the chromo-electric field 
self consistently, as a product of color currents, which a affected also 
by the field. Our effective fields are external with respect to colliding 
partons, that why we can use the expression (\ref{eq22}) for the field energy.
On the other hand, if we want to satisfy the conservation laws, we must  
generate our effective fields in the collision transferring energy from 
matter to field.
It's possible to define new conserved quantities based on 
eq. (\ref{eq1}).  
Using the definition of $F^{\mu\nu}$, eq.
(\ref{eq7}), we can rewrite eq. (\ref{eq1}) as
\beq
\partial_\mu T^{\mu\nu} =\sum_i F_i^{\mu\nu} q_i N_{i, \mu}= 
\sum_i q_i \left[\partial^\mu \left(A_i^\nu N_{i, \mu}\right) -
A_i^\nu \partial^\mu N_{i, \mu}
-\partial^\nu \left(A_i^\mu N_{i, \mu}\right)
+A_i^\mu \partial^\nu N_{i, \mu}
\right] \ .
\eeq{eq25}

The solutions for $N_1^-$ and $N_2^+$, eq. (\ref{eq21}), show that the
second term vanishes. The fourth term is a vector
$-(A_1^-\partial_-N_1^+,A_2^+\partial_+N_2^-)$ in lightcone coordinates.
So, if we impose the conditions
\beq
A_1^-=0\ , \quad A_2^+=0
\eeq{apm}
we can define a new energy-momentum
tensor $\tilde{T}^{\mu\nu}$, such that
\beq
\partial_\mu \tilde{T}^{\mu\nu}=0 \ ,
\eeq{eq26}
\beq
\tilde{T}^{\mu\nu}=\sum_i \tilde{T_i}^{\mu\nu}+T^{\mu\nu}_{pert}=
\sum_i \left(T_i^{\mu\nu} - q_i A_i^\nu  N_i^\mu +
g^{\mu\nu} q_i A_i^\alpha N_{i \alpha} \right)+B g^{\mu\nu}
\eeq{eq27}

To satisfy the above choice of the fields, (\ref{sch}), and imposed
conditions (\ref{apm}) together with
the Lorentz gauge condition, 
$\partial^0 A_i^0 - \partial^3 A_i^3 = 0$ or
$\partial^- A_i^- - \partial^+ A_i^+ = 0$, we take the 
vector potentials in the following form:
\beq
\begin{array}{ll}
A_1^- = 0, & A_1^+ = - \sigma_1 x^+ = - \sigma x^+, \\
A_2^- = \sigma_2 x^- = \sigma x^- ,& A_2^+ = 0 \ .
\end{array}
\eeq{eq24}
Notice that the above choice differs from the one which was initially 
proposed in Refs. \cite{MCS00,CAM00,MCS00-1}, what causes the changes 
in the expressions related to field creation in this section - eqs. 
(\ref{eq27a},\ref{eq42},\ref{eq43}) - but will not affect the analytic 
solution of the model - eqs. 
(\ref{eq54}-\ref{eq60}). 

In our
calculations we used the parameterization:
\beq
\sigma=A\left(\frac{\varepsilon_0}{m}\right)^2 n_0
\sqrt{l_1l_2} \ ,
\eeq{eq44}
where $m$ is the nucleon mass, and 
$l_1$ and $l_2$ are the initial streak lengths (see Fig.
\ref{fig1}). We are working in
the system, 
where $\ \hbar=c=1$, so $\sigma$ has a dimension of 
$length^{-2}=energy/length$. 
The typical values of dimensionless parameter 
$A$ are around $0.06-0.08$. 
Notice, that there is only one free parameter in
parameterization (\ref{eq44}).  The typical values of $\sigma$ are $4-10\
GeV/fm$ for $\varepsilon_0=65 \ GeV$ per nucleon, and
$\sigma \approx 6-15\ GeV/fm$ for $\varepsilon_0=100 \ GeV$ per nucleon.  
These values are
consistent with the energy density in non-hadronized strings, or "latent
energy density" which is on the average $9\ GeV/fm^3$
\cite{ASC91pl,ASC91prl,ACS92}.

Using the exact definition of $A_i^\mu$, eqs. (\ref{eq24}), eqs.
(\ref{eq14},\ref{eq17a},\ref{eq22},\ref{sch},\ref{eq27}) 
and transformation matrixes from \ref{app3}
we obtain
$$
\tilde{T}^{\mu\nu}=\left(
\begin{array}{cc}
\tilde{T}^{++} & \tilde{T}^{+-} \\
\tilde{T}^{-+} & \tilde{T}^{--}
\end{array}\right)=
\frac{1}{2}\left(
\begin{array}{cc}
h_{1+}e^{2y_1} & h_{1-} \\
h_{1-} & h_{1+}e^{-2y_1}
\end{array}\right)
$$
\beq
+\frac{1}{2}\left(
\begin{array}{cc}
h_{2+}e^{2y_2} & h_{2-} \\ 
h_{2-} & h_{2+}e^{-2y_2} 
\end{array}\right)
+\frac{1}{2}\left(
\begin{array}{cc}
\sigma^2 & 2B \\
2B & \sigma^2
\end{array}\right) 
\eeq{eq27a}
$$
+\left(
\begin{array}{cc}
-\sigma x^+ N_1^+ & 0\\
\sigma x^+ N_1^- & 0
\end{array}\right)
+\left(
\begin{array}{cc}
0&\sigma x^- N_2^+\\
0&-\sigma x^- N_2^-
\end{array}\right)
$$
Notice that perturbative term, $B$, and free field energy, 
$\frac{\sigma^2}{2}$, cover all the interacting volume, while 
energy densities of matter and baryon currents are separated in space.
We also want to stress factor $1/2$ in front of all terms in 
$\tilde{T}^{\mu\nu}$ (it has been canceled by $2$ near $\sigma$ in 
last two terms) - this factor was missed in Refs. \cite{MCS00,CAM00,MCS00-1} 
as well as in \cite{GC86}, although it does not affect the result since 
equations of motion, $\partial_\mu \tilde{T}^{\mu\nu}=0$, can be 
multiplied by any coefficient. The reason for it is a form of transformation 
matrixes between $(t,z)$ and $(+,-)$ coordinates, which are presented in 
\ref{app3}.

Now the new conserved quantities are
\beq
Q_0=\int \tilde{T}^{00} dV = \dlt x\dlt y 
\int \tilde{T}^{00} dz
\ ,
\eeq{eq28}
\beq
Q_3=\int \tilde{T}^{03} dV = \dlt x \dlt y 
\int \tilde{T}^{03} dz
\ ,
\eeq{eq29}
where the volume integral runs over
the lengths of the both streaks and 
$\dlt x\dlt y$ is the cross section of the streaks. Notice that in the 
absence of the fields, before string creation and after string decay, the
$(Q^0,Q^3)$ come back to $(P^0,P^3)$ - components of the 4-momenta of 
the system.

We can rewrite energy-momentum tensor in $(t,z)$ coordinates:
$$
\tilde{T}^{\mu\nu}=\left(
\begin{array}{cc}
\tilde{T}^{00} & \tilde{T}^{03} \\
\tilde{T}^{30} & \tilde{T}^{33}
\end{array}\right)=
\left(
\begin{array}{cc}
(e_1+P_1)\cosh^2 y_1 - P_1& (e_1+P_1)\cosh y_1 \sinh y_1\\
(e_1+P_1)\cosh y_1 \sinh y_1& (e_1+P_1)\sinh^2 y_1 + P_1
\end{array}\right)
$$
\beq
+\left(
\begin{array}{cc}
(e_2+P_2)\cosh^2 y_2 - P_2& (e_2+P_2)\cosh y_2 \sinh y_2\\
(e_2+P_2)\cosh y_2 \sinh y_1& (e_2+P_2)\sinh^2 y_2 + P_2
\end{array}\right)
+\left(
\begin{array}{cc}
\frac{\sigma^2}{2}+B & 0 \\
0& \frac{\sigma^2}{2}-B
\end{array}\right) 
\eeq{eq27d}
$$
+\frac{\sigma x^+}{2}\left(
\begin{array}{cc}
 N_1^- - N_1^+ & N_1^- - N_1^+  \\
-(N_1^- + N_1^+) & -(N_1^- + N_1^+) 
\end{array}\right)
+\frac{\sigma x^-}{2}\left(
\begin{array}{cc}
 N_2^+ - N_2^- & -(N_2^- - N_2^-)  \\
 N_2^+ + N_2^- & -(N_2^+ + N_2^-) 
\end{array}\right)
$$
Based on conservation of $Q_0,\ Q_3$ we can calculate energy 
densities, $e_1(t_0),\ e_2(t_0)$, at the moment $t=t_0$, 
when the string with tension
$\sigma$ is created. 
These new quantities are used as initial conditions
for our differential eqs. (\ref{eq1}, \ref{eq2}). As shown in 
\ref{app1} --
\beq
e_1(t_0)=
\frac{n_0 m}{1{+}c_0^2}-
\frac{{\sigma^2\over 2} + B}{ \left(\frac{\varepsilon_0}{m}\right)^2
 (1{+}c_0^2)}
\frac{l_1{+}l_2}{2l_1}-
\frac{\sigma n_0 e^{y_0}}{4\left(\frac{\varepsilon_0}{m}\right)^2
(1{+}c_0^2)}l_1\ ,
\eeq{eq42}
\beq
e_2(t_0)=
\frac{n_0 m}{1{+}c_0^2}-
\frac{{\sigma^2\over 2} + B}{ \left(\frac{\varepsilon_0}{m}\right)^2
 (1{+}c_0^2)}
\frac{l_1{+}l_2}{2l_2}-
\frac{\sigma n_0 e^{y_0}}{4\left(\frac{\varepsilon_0}{m}\right)^2
(1{+}c_0^2)}l_2\ .
\eeq{eq43}
Here the $e_i(t_0)$ is a proper energy density at the time $t_0$,
$\varepsilon_0$ is the initial energy per nucleon.
We assumed transparency,  i.e., that complete penetration happened so fast, 
that the fields, created during this time, did not have time to stop partons.
So, the rapidities are $y_{1(2)}(t_0)=-y_0(y_0)$ correspondingly, and 
the proper baryon densities did not change.

For $x^\pm>x_0$ we should solve eqs. (\ref{eq26}), with boundary conditions
\beq
\begin{array}{ll}
N_1^\pm (x^-=x_0)= n_0 e^{\mp y_0} &
N_2^\pm (x^+=x_0)= n_0 e^{\pm y_0} \\ & \\
h_{1+}(x^-=x_0)=e_1(t_0)(1+c_0^2) &
h_{2+} (x^+=x_0)=e_2(t_0)(1+c_0^2) \\ & \\
y_1(x^-=x_0)=-y_0 & y_2(x^+=x_0)=y_0 \\ & \\
\sigma_1(x^-=x_0)=\sigma & \sigma_2(x^+=x_0)=\sigma\\ & \\
q_1(x^-=x_0)=1 & q_2(x^+=x_0)=-1\ ,
\end{array}
\eeq{eq45}
where $x_0=2t_0-|z(0)|$ defines the string creation surface $t=t_0$, 
for parton or cell element in the position $z=z(0)$ at the time $t=0$.

Let us present the complete analytical solution 
in the following form (for detailed calculations see \ref{app2})
\beq
e^{(-)^{i+1}2y_i}=-\frac{d_i}{b_i}+
\left(\frac{d_i}{b_i}+e^{-2y_0}\right)
\left(1-\frac{x^i-x_0}{\tau_i}\right)^{-\frac{b_i}{\alpha a_j}} \ ,
\eeq{eq54}
\beq
h_{i+}=e^{(-)^{i+1}2y_i}e_i(t_0)(1+c_0^2)e^{2y_0}
\left(1-\frac{x^i-x_0}{\tau_i}\right)\ ,
\eeq{eq55}
\beq
 n_i= n_0e^{y_0}e^{(-)^{i+1}y_i} \ ,
\eeq{eq56}
where $x^1=x^-,\ x^2=x^+$, $i,j=1,2\ ,\ i\not=j$, \ \ \
and the notations are from \ref{app2} 
(\ref{eq52},\ref{eq61},\ref{eq57.b}-\ref{eq58}).

Then the trajectories of partons (or cell elements)
for both nuclei are given by:
\beq
\begin{array}{c}
x_1^+(x^-)=|z(0)|+\int_{x_0}^{x^-}dx\ e^{2y_1(x)}=\nonumber 
\\ \\ |z_0|-\frac{d_1}{b_1}(x^--x_0)+
\left(\frac{d_1}{b_1}+e^{-2y_0}\right)
\tau_1\frac{\alpha a_2}{2\sigma n_0e^{y_0}}
\left[
\left(1-\frac{x^--x_0}{\tau_1}\right)^{-\frac{2\sigma n_0e^{y_0}}
{\alpha a_2}}
- 1\right] \ ,\nonumber
\end{array}
\eeq{eq59}
\beq
\begin{array}{c}
x_2^-(x^+)=|z(0)|+\int_{x_0}^{x^+}dx\ e^{-2y_2(x)}=\nonumber 
\\ \\ |z(0)|-\frac{d_2}{b_2}(x^+-x_0)+\left(\frac{d_2}{b_2}+
e^{-2y_0}\right)
\tau_2\frac{\alpha a_1}{2\sigma n_0e^{y_0}}
\left[
\left(1-\frac{x^+-x_0}{\tau_2}\right)^{-\frac{2\sigma n_0e^{y_0}}
{\alpha a_1}}
- 1\right] \ ,\nonumber \end{array} 
\eeq{eq60}
for parton or cell element in the position $z=z(0)$ at the time $t=0$.

\section{Recreation of matter}

\label{four}

If we let partons (or cell domains) evolve according to the above 
trajectories, eqs. (\ref{eq59}, \ref{eq60}),
they will keep going in the initial direction up to
the time $t=t_{i,turn}$, then they will turn and go backwards until the
two streaks again penetrate through each other and a new oscillation will
start. Such a motion is analogous to the "Yo-Yo" motion in the string
models.  Of course, it is difficult to believe that such a process would
really happen in heavy ion collisions, because of string decays,
string-string interactions, interaction between streaks and other reasons,
which would be difficult to take into account.  To be realistic we should
stop the motion described by eqs. (\ref{eq59}, \ref{eq60}) at some moment
before the projectile and target cross again.

We assume that the final result of collisions of two streaks, after
stopping the string's expansion and after its decay, is one streak of the
length $\dlt l_f$ with homogeneous energy density distribution, $e_f$, and
baryon charge distribution, $ n_f$, moving like one object with rapidity
$y_f$. We assume that this is due to string-string interactions and string
decays. As was mentioned above the typical values of the string
tension, $\sigma$, are of the order of $10\ GeV/fm$, and these may be
treated as several parallel strings. The string-string interaction will
produce a kind of "string rope" between our two streaks, which is
responsible for final energy density and baryon charge 
distributions. For simplicity we assume homogeneous baryon charge
distribution.
Notice, that in this way, after the decay of our "string rope"
charges do not remain at the ends of the final streak, as it would be if we
assume full transparency.
The real situation may be more complicated: when the energy accumulated in 
the strong color fields will be finally released in a production of 
$q\bar{q}$ pairs and gluons, this process may noticeably change 
composition of matter as compared to the chemical equilibrium case
\cite{KM85}. Therefore, matter created after the mutual stopping 
of interpenetrating streaks can not, in general, be described 
by the equilibrium EoS. 
The homogeneous distributions are the simplest assumptions, which may be
modified later based on experimental data. 
Its advantage is a simple expression
for $e_f,\  n_f,\ y_f$. 
The first experimental results from RHIC do not show transparency, rather 
most particle multiplicities as well as the elliptic flow show strong
stopping and a peak around mid rapidity \cite{QM01}. Furthermore,
we describe the initial state, which is not directly observable in experiments,
and a flat initial rapidity distribution  may end up in both
a forward-backward peaked and a centrally peaked distributions
depending on several other circumstances. 

The final energy density, baryon density and rapidity, $e_f, \  n_f$ and
$y_f$, should be determined from conservation laws. The 
assumptions we made above oversimplify the situation, and do not 
allow us to satisfy exactly all conservation laws. The 
reason for this is well known and has been discussed in the 
Refs. \cite{FO2,FO2a,FO3,FO3a,FO3b}:  
two possible definitions of the flow, Eckart's and Landau's definition.
If we are following the energy
flow, we satisfy exactly the energy and momentum conservation, but 
violate the net baryon current conservation. (Otherwise, if we were to choose
baryon flow, we would
violate the energy-momentum conservation.)      

The exact conservation of the energy and momentum gives for the final rapidity:
\beq
\cosh^2 y_{f,L} = \frac{(M^2(1+c_0^2)+2c_0^2v_0^2)+
\sqrt{(M^2(1+c_0^2)+2c_0^2v_0^2)^2+4c_0^4v_0^2(M^2-v_0^2)}}{2(1+c_0^2)
(M^2-v_0^2)} \ ,
\eeq{eq66}
where we neglected $B \dlt l_f$ next to $Q_0/\dlt x\dlt y$ 
and introduced the 
notation
$M=(l_2+l_1)/(l_2-l_1)$, $v_0=\tanh y_0$ is the initial velocity.
(The exact conservation of the baryon four-current would give:  
$
\tanh y_{f,E} = v_0/M \ $, $\rightarrow \  
$
$
\cosh^2 y_{f,E} = M^2/(M^2-v_0^2) \ 
$).

It is interesting to analyze these equations, 
as functions
of $l_1$ and $l_2$.  If $l_1$ or $l_2\ \rightarrow \ 0$ then
$M^2\rightarrow1$, and $|y_{f,E}|\rightarrow y_0$. To calculate this limit
for $|y_{f,L}|$ we should put $c_0^2=0$, since we do not have collisions 
and, consequently, do not create QGP, thus $|y_{f,L}|\rightarrow y_0$.   
So there is no stopping 
as expected, because there is
no reason to stop.  
If $l_1\rightarrow l_2 \quad M^2\rightarrow \infty$ and 
both expressions give
$y_{f,E,L}\rightarrow 0$, i.e., complete stopping. 
So, we see that Landau's and Eckart's expressions
behave similarly, have the same limits for minimal and maximal stopping. 
  
For the following part of this work we choose Landau's convention, 
$y_{f}=y_{f,L}$, which is justifiable for RHIC and SPS energies,
where the evolution of matter is not dominated by the net baryon
charge, unlike at lower energies where the baryon mass is still
dominant and pair creation is of little importance.

In this case the expressions for the $e_f$ and $ n_f$ are:
\beq
e_f=\frac{{Q_0\over\dlt x \dlt y}}{((1+c_0^2)\cosh^2 y_f 
- c_0^2)
\dlt l_f} \ ,
\eeq{eq68}
\beq
 n_f=\frac{ n_0(l_1+l_2)}{\dlt l_f \cosh y_f} \ .
\eeq{eq69}

\begin{figure}[htb]
\insertplotshot{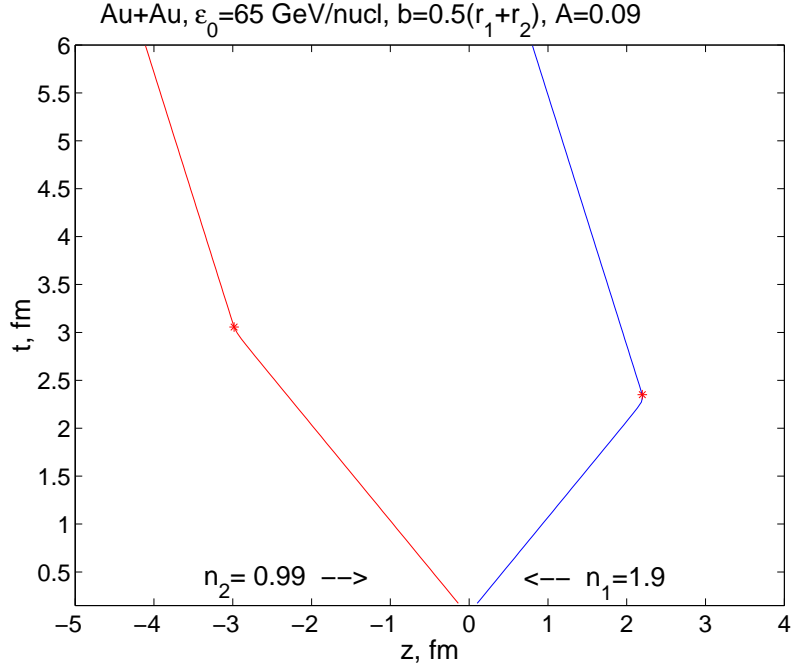} 
\vspace{0.6cm}
\caption[]{The typical trajectory of the ends of two 
initial streaks, corresponding to numbers of nucleons, $n_1$ and $n_2$. 
Stars denote the points, where $y_i=y_f$. From $t=t_0$ till those 
streak ends keep going according to eqs. (\ref{eq59}, \ref{eq60}). 
Later the final streak starts to move 
like one object with rapidity, $y_f$, eq. (\ref{eq66})
in CRF.}

\label{fig2}
\end{figure}

The typical trajectory of the streak-ends is presented in Fig.  \ref{fig2}.
 From $t=t_0$ they move 
according to eqs. (\ref{eq59}, \ref{eq60}) until they reach the rapidity 
$y_i=y_f$.
Later the final streak starts to move
like one object with uniform rapidity, $y_f$, until we reach the time 
when the fluid dynamical calculation starts.

The time and position of final streak formation
 can be found from the condition:
\beq
y_i=y_f \ ,
\eeq{eq62}
which gives for the $i$th nucleus ($x_1=x^-,\ x_2=x^+$)
\beq
x_{i,\ final}=x_0+\tau_i
\left[1-\left(\frac{\frac{d_i}{b_i}+e^{-2y_0}}
{\frac{d_i}{b_i}+e^{(-)^{i+1}2y_f}}
\right)^{\frac{\alpha a_j}{b_i}}\right] \ .
\eeq{eq63}

\begin{figure}[htb]
\insertplot{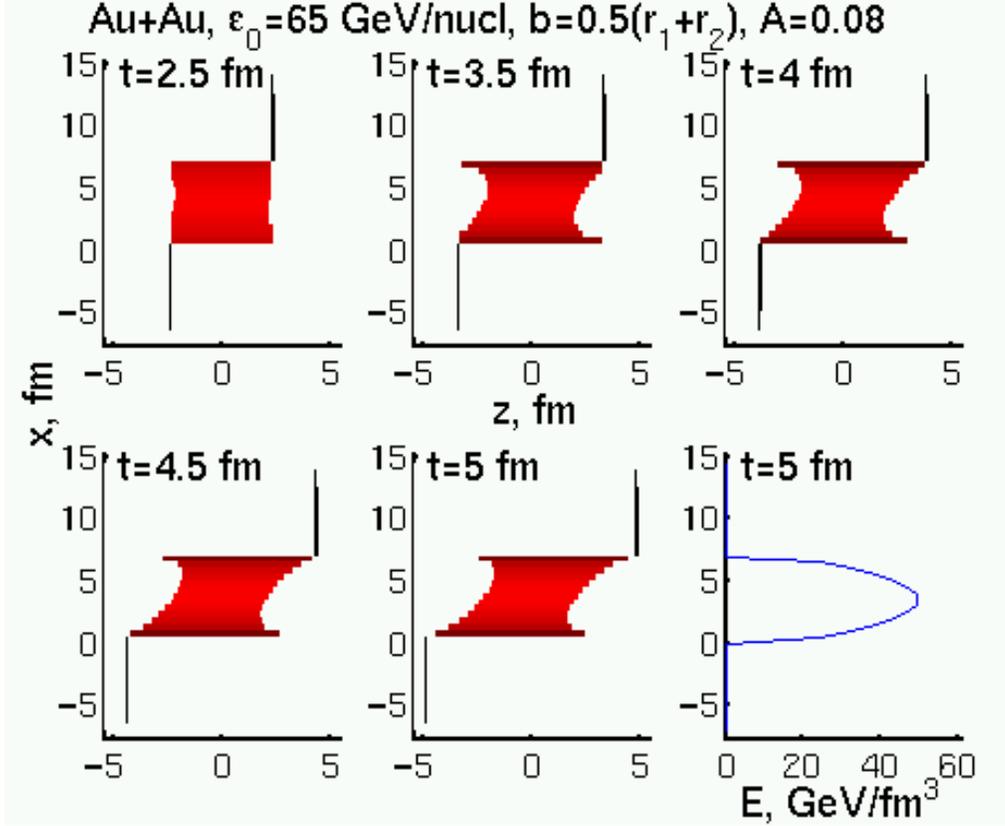}
\vspace{0.6cm}
\caption[]{The Au+Au collision at $\varepsilon_0=65\ GeV/nucl$, 
$b=0.5(r_1+r_2)$, $A=0.08$ (parameter $A$ was introduced in 
eq. (\ref{eq44})),
 $y=0$ (ZX plane through 
the centers of nuclei), $E=T^{00}$, laboratory frame. 
We note that 
the final shape of the QGP volume is a tilted disk $\approx 45^0$, 
and the direction of the fastest expansion will deviate from both 
the beam axis and the usual transverse flow direction and might be a 
reason for the third flow component, as argued in \cite{CR}.}

\label{ev11}
\end{figure}

\section{Initial conditions for hydrodynamic calculations}

\label{five}

In this section we present the results of our calculations.  We are
interested in the shape of the QGP formed when string expansions stop and
their matter is locally equilibrated.  This will be the initial state for
further hydrodynamic calculations.  The time, $\tau$, at which we assume
the system to reach overall 
local equilibrium and to start hydrodynamic description, is a
second (after $A$) free parameter of our model. Of course, $\tau$ should
be larger than the time of final streak formation, at least in the central
most hot and dense region. For the peripheral streaks the string tension
is low, and the transparency is large, but peripheral matter does not play
a leading role in further hydrodynamic expansion. So, to have a
homogeneous output for each streak-streak collision, we will also build
the final streaks ($y_f,\  n_f,\ e_f$) for peripheral streak-streak
collisions, with lengths, $\dlt l_f$, corresponding the lengths of the
interacting region at the moment $t=\tau$, even if the final rapidity,
$y_f$, was not yet achieved for this particular collision.

\begin{figure}[htb]
\insertplot{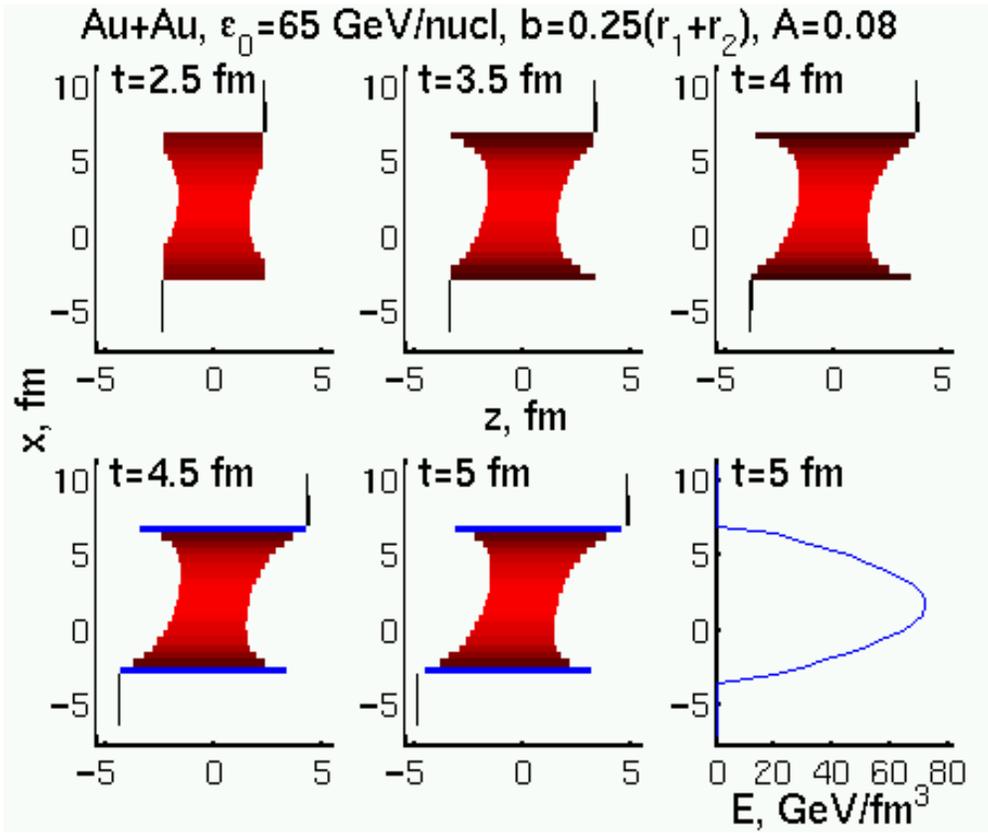}
\vspace{0.6cm}
\caption[]{The same as Fig. \ref{ev11}, but $b=0.25(r_1+r_2)$. We see that 
for more central collisions the energy density is much larger. 
The QGP volume 
has a shape of tilted disk and may produce a third flow component 
\cite{CR}. }

\label{ev12}
\end{figure}

\begin{figure}[htb]
\insertplot{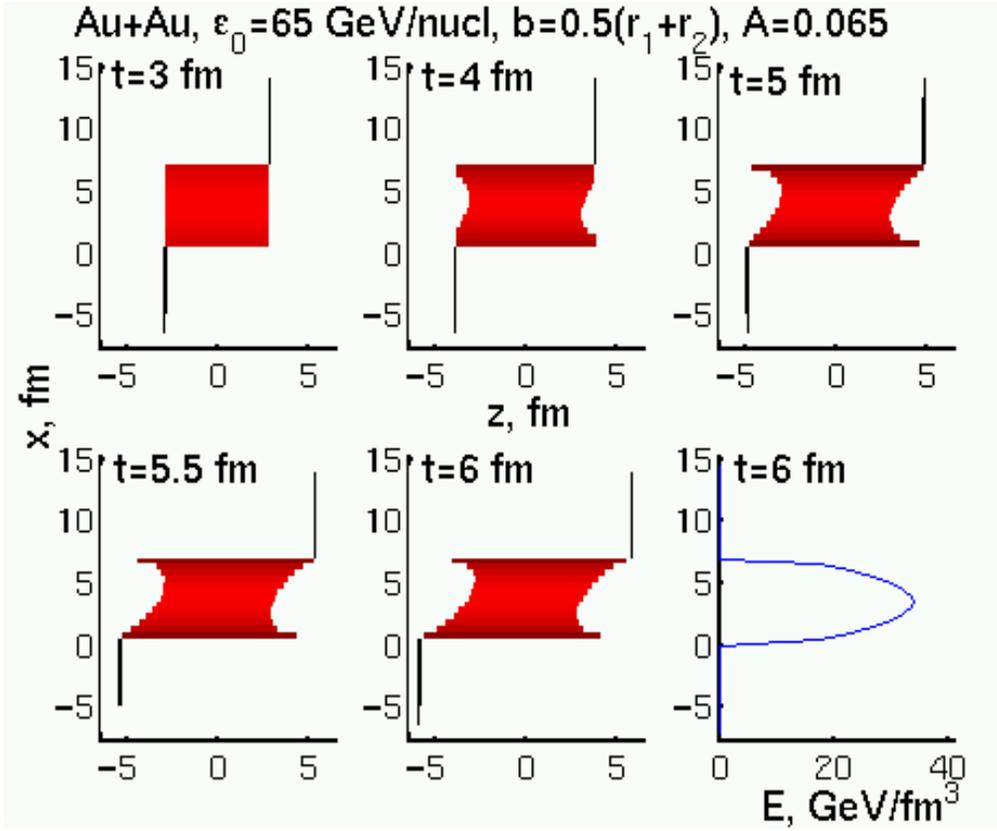}
\vspace{0.6cm}
\caption[]{The same as Fig. \ref{ev11}, but $A=0.065$.  The energy density
is smaller, but the QGP volume has a similar shape of a tilted disk $\approx
45^0$ and may produce a third flow component \cite{CR}. We start plotting
our results later than in Fig. \ref{ev12}, because for smaller $\sigma$
the deceleration is smaller, and, so, the final streak is formed later.}

\label{ev21}
\end{figure}

\begin{figure}[htb]
\insertplot{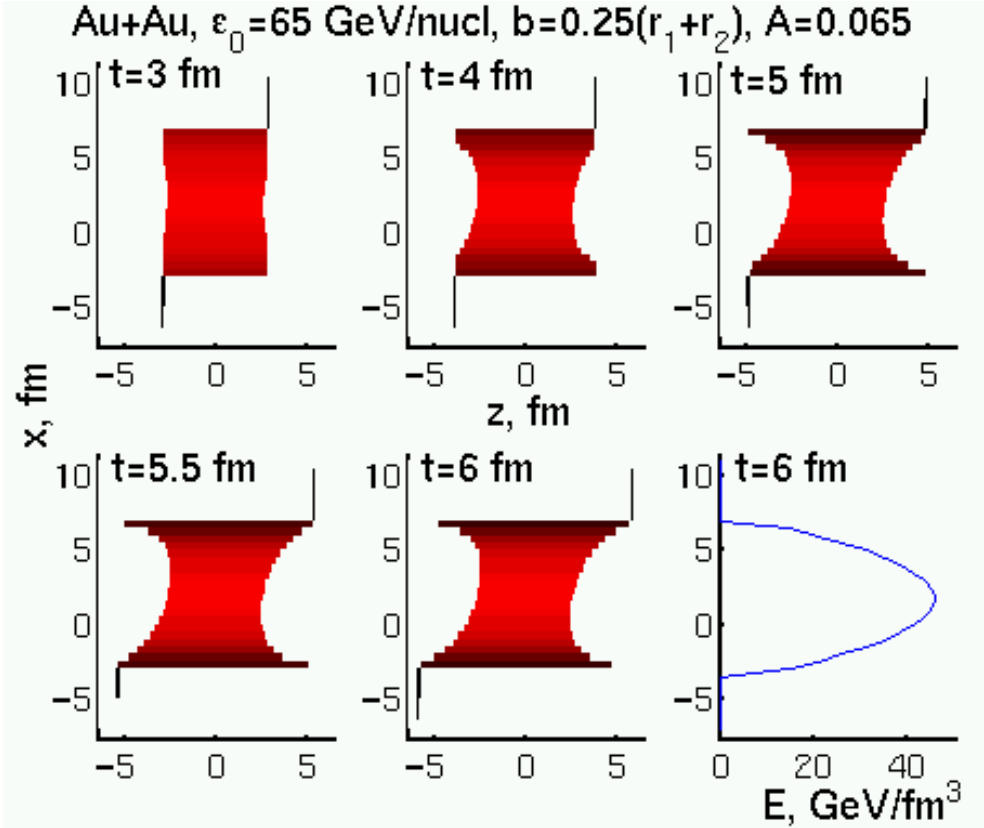}
\vspace{0.6cm}
\caption[]{The same as Fig. \ref{ev12}, but $A=0.065$. We see that
for more central collisions -- compared to Fig. \ref{ev21} -- the energy
density is much larger, but it is smaller than in Fig.  \ref{ev12},
because of smaller stopping. The QGP volume has a shape of tilted disk and
may produce a third flow component \cite{CR}. }

\label{ev22}
\end{figure}

We may see in Figs. \ref{ev11}, \ref{ev12}, \ref{ev21}, \ref{ev22}
that finally a QGP forms a tilted disk for
$b\not =0$. Thus, the direction of fastest expansion, the same as largest
pressure gradient, will be in the reaction plane, but will deviate from
both the beam axis and the usual transverse flow direction.  So, the new
flow component, called "antiflow" or "third flow component" \cite{CR},
will appear in addition to the usual transverse flow component in the
reaction plane.  With increasing beam energy the usual transverse flow is
getting weaker, while this new flow component is strengthened. The mutual
effect of the usual directed transverse flow and this new "antiflow" or
"third flow component" contribute to an enhanced emission in the reaction
plane.  This was actually observed and studied earlier. One should also
mention that both the standard transverse flow and new "antiflow"
contribute to "elliptic flow".

The last subplots in the Figs. \ref{ev11}, \ref{ev12}, 
\ref{ev21}, \ref{ev22}
present the energy density distribution in the laboratory frame, 
$E_{max}\approx 50-90\ GeV/fm^3$ for $b=0$.
It seems to be bigger than what one can expect from estimation based on
the Bjorken model. One should, nevertheless, keep in mind that our "fireball"
is not homogeneous in $xy$ plane. The average energy density for the 
equivalent homogeneous "fireball" would be lower -
$<E>=22-29\ GeV/fm^3$. 
Other hydrodynamical models had to use similarly high 
initial energy density to reproduce the observed flow, e.g. in \cite{P_qm} 
$\epsilon_0=23\ GeV/fm^3$ has been used.

\section{Conclusions}

\label{six}

Based on earlier Coherent Yang-Mills field theoretical models and
introducing effective parameters, based on Monte-Carlo string cascade and
parton cascade model results, a simplified model is introduced to describe
the pre fluid dynamical stages of heavy ion collisions at the highest SPS
energies and above.  The model predicts limited transparency for massive
heavy ions.

Contrary to earlier expectations --- based on standard string tensions of
$1\ GeV/fm$ which lead to the Bjorken model type of initial state ---
effective string tensions are introduced for collisions of massive heavy
ions.  The increased string tension is a consequence of collective effects
related to QGP formation.  These collective effects in central and semi
central collisions lead to an effective string tension of the order of $10\
GeV/fm$ and consequently cause much less transparency than earlier
estimates. The resulting initial locally equilibrated state of matter in
semi central collisions takes a rather unusual form, which can be then
identified by the asymmetry of the caused collective flow.  Our prediction
is that this special initial state may be the cause of the recently
identified "antiflow" or "third flow component". 

Detailed fluid dynamical calculations as well as flow experiments at
semi central impact parameters for massive heavy ions are needed 
at SPS and RHIC energies to connect the predicted special initial state
with observables.

\section*{Acknowledgment}

One of the authors (D.D.S.) thanks the 
support of the Bergen Computational Physics Laboratory in the
framework of the European Community - Access to Research Infrastructure 
action of the Improving Human Potential Programme and 
the Humboldt Foundation. 

\appendix

\begin{figure}[htb]
\insertplot{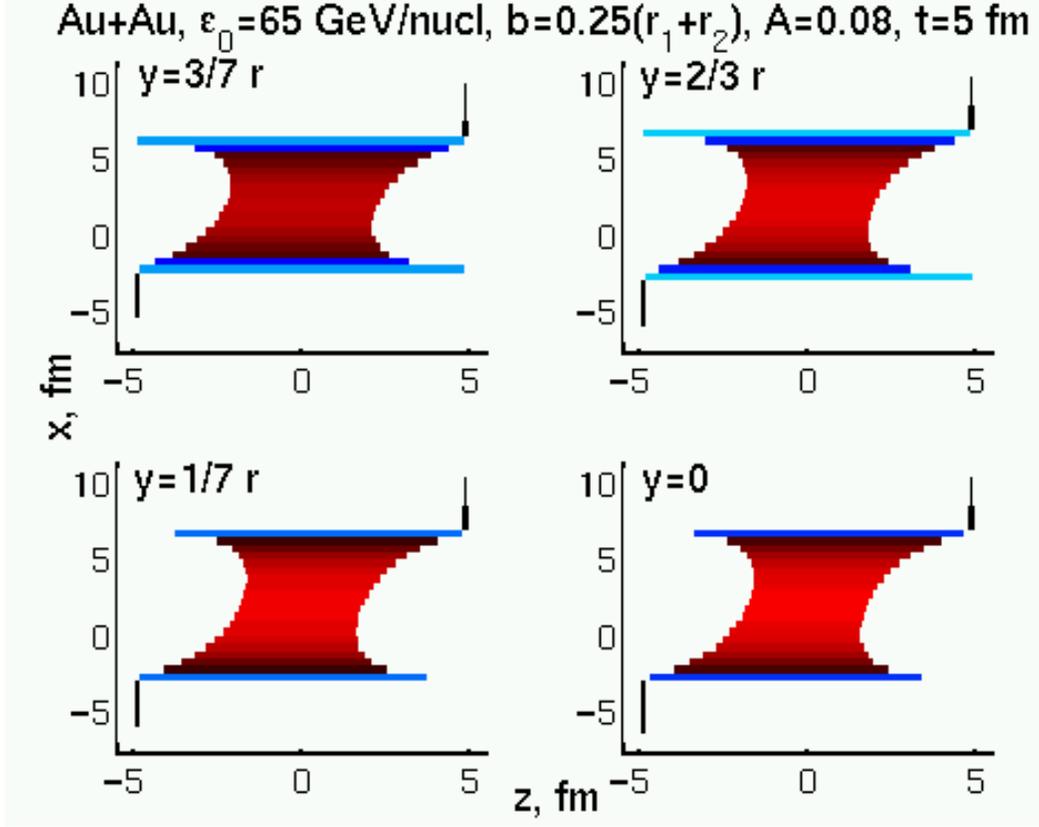}
\vspace{0.6cm}
\caption[]{The Au+Au collision at $\varepsilon_0=65\ GeV/nucl$, 
$t=5\ fm$, $A=0.08$ (parameter $A$ was introduced in
(\ref{eq44})), $b=0.25(r_1+r_2)$ 
(in our case $r_1=r_2=r$).  We see that the more central plane we 
look at, the more
nucleons take part in the streak-streak collisions, and therefore the more
energetic and compact QGP is formed.}

\label{ydist1}
\end{figure}

\section{$\!\!\!\!\!\!$: Initial conditions after string creation}
\label{app1}
Our conserved quantities are (\ref{eq28},\ref{eq29})
\beq
Q_0=\int \tilde{T}^{00} dV = \dlt x\dlt y 
\int \tilde{T}^{00} dz
\ ,
\eeq{eq28a}
\beq
Q_3=\int \tilde{T}^{03} dV = \dlt x \dlt y 
\int \tilde{T}^{03} dz
\ ,
\eeq{eq29a}
$\tilde{T}^{00}$ and $\tilde{T}^{03}$ are given by eq. (\ref{eq27d}).
Before string creation the initial values of the modified 
energy-momentum tensor, 
$\tilde{T}^{\mu\nu}$, are
\beq
\tilde{T}_1^{00}=\tilde{T}_2^{00}=e_0\cosh^2 y_0=
\left(\frac{\varepsilon_0}{m}\right)^2 n_0 m \ ,
\eeq{eq34.0}
\beq
\tilde{T}_2^{03}=-\tilde{T}_1^{03}=e_0\tanh y_0 \cosh^2 y_0=
\left(\frac{\varepsilon_0}{m}\right)^2 n_0 m v_0\,
\eeq{eq34.3}
where $m$ is the nucleon mass, $\varepsilon_0$ is the initial 
energy per nucleon, and 
we have used 
$\cosh^2 y_0=\gamma_0^2=\left(\frac{\varepsilon_0}{m}\right)^2$.
$v_0=\tanh y_0$ 
is the initial velocity, $v_0=1$ is a good approximation for 
ultra-relativistic heavy ion collisions. 
So,
\beq
Q_0=\dlt x \dlt y \left(\frac{\varepsilon_0}{m}\right)^2 n_0 m
(l_1+l_2) \ ,
\eeq{eq32}
\beq
Q_3=\dlt x \dlt y \left(\frac{\varepsilon_0}{m}\right)^2 n_0 m
(l_2-l_1) v_0 \ ,
\eeq{eq33}
where $l_1$ and $l_2$ are the initial lengths 
of streaks (see Fig \ref{fig1}), $\dlt x$, $\dlt y$ are the grid sizes 
in $x$ and $y$ directions.

After string creation
$$
\tilde{T}^{00}=e_1 \cosh^2 y_1 +c_0^2e_1 \sinh^2 y_1
+e_2 \cosh^2 y_2 +c_0^2e_2\sinh^2 y_2
$$
\beq
+{1\over 2}\sigma^2+B+
\frac{\sigma x^+}{2} n_0 e^{y_0} 
\left(1-e^{2y_1}\right) 
+\frac{\sigma x^-}{2} n_0 e^{y_0} 
\left(1-e^{-2y_2}\right) \ ,
\eeq{eq36}
$$
\tilde{T}^{03}=e_1(1+c_0^2) \cosh y_1\sinh y_1
+e_2(1+c_0^2) \cosh y_2\sinh y_2
$$
\beq
-\frac{\sigma x^+}{2} n_0 e^{y_0} 
\left(1+e^{2y_1}\right) 
+\frac{\sigma x^-}{2} n_0 e^{y_0} 
\left(1+e^{-2y_2}\right) \ ,
\eeq{eq38}

\begin{figure}[htb]
\insertplot{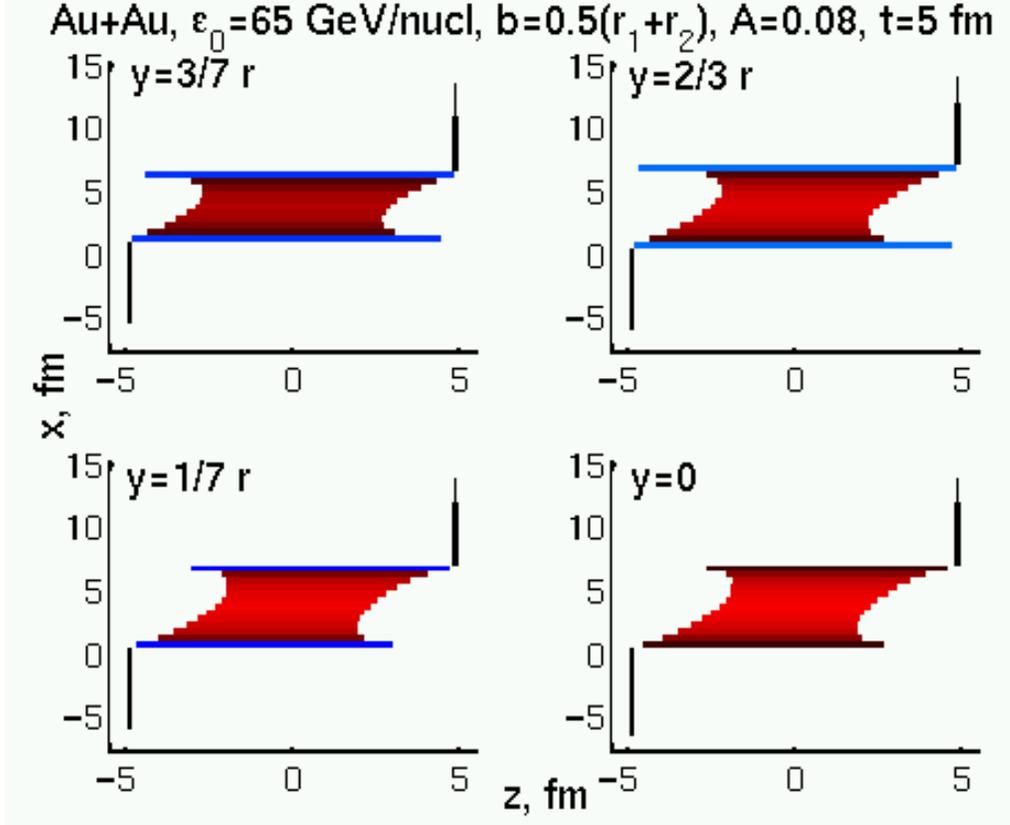}
\vspace{0.6cm}
\caption[]{The same as Fig. \ref{ydist1}, but $b=0.5(r_1+r_2)$. The stopping 
is smaller, consequently the QGP volume is less dense and less compact.}

\label{ydist2}
\end{figure}

At the point of complete penetration of streaks, $t=t_0=(l_1+l_2)/2$ 
(see Fig \ref{fig1}), we introduced energy densities $e_1(t_0)$ and 
$e_2(t_0)$. 
We assumed transparency,  i.e., that complete penetration happened so fast, 
that field, itself created during this time, did not have time to stop partons.
So, the rapidities $y_{1(2)}(t_0)=-y_0(y_0)$ correspondingly, and 
the proper baryon densities did not change, and, thus, the baryon current 
conserved automatically. This assumption differs from how it was done
in Refs. \cite{MCS00,CAM00,MCS00-1}, but they seems to be more physical, 
and do not change final results very much.  
Terms proportional to $e^{2y_1}=e^{-2y_2}=e^{-2y_0}\ll 1$ can be neglected.
Then the energy and momentum conservation laws can be written in the form:
\beq
\frac{Q_0}{\dlt x \dlt y}=
\left[(1+c_0^2) \cosh^2 y_0 - c_0^2\right]\left(e_1(t_0)l_1+
e_2(t_0)l_2\right)
+\left(\frac{\sigma^2}{2}+B\right)(l_1+l_2)+
\frac{\sigma  n_0 e^{y_0}}{4}\left(l_1^2+l_2^2\right)\ ,
\eeq{eq40}
\beq
\frac{Q_3}{\dlt x \dlt y}=
\left[(1+c_0^2) \cosh^2 y_0\right]\left(-e_1(t_0)l_1+
e_2(t_0)l_2\right)
-\frac{\sigma  n_0 e^{y_0}}{4}\left(l_1^2-l_2^2\right)
\ .
\eeq{eq41}

We neglect $c_0^2$ close to $(1+c_0^2) \cosh^2 y_0$ in eq. \ref{eq40},
then eqs. (\ref{eq40}, \ref{eq41}) may be solved 
\beq
e_1(t_0)=
\frac{n_0 m}{1{+}c_0^2}-
\frac{{\sigma^2\over 2} + B}{ \left(\frac{\varepsilon_0}{m}\right)^2
 (1{+}c_0^2)}
\frac{l_1{+}l_2}{2l_1}-
\frac{\sigma n_0 e^{y_0}}{4\left(\frac{\varepsilon_0}{m}\right)^2
(1{+}c_0^2)}l_1\ ,
\eeq{eq42a}
\beq
e_2(t_0)=
\frac{n_0 m}{1{+}c_0^2}-
\frac{{\sigma^2\over 2} + B}{ \left(\frac{\varepsilon_0}{m}\right)^2
 (1{+}c_0^2)}
\frac{l_1{+}l_2}{2l_2}-
\frac{\sigma n_0 e^{y_0}}{4\left(\frac{\varepsilon_0}{m}\right)^2
(1{+}c_0^2)}l_2\ .
\eeq{eq43a}

\begin{figure}[htb]
\insertplot{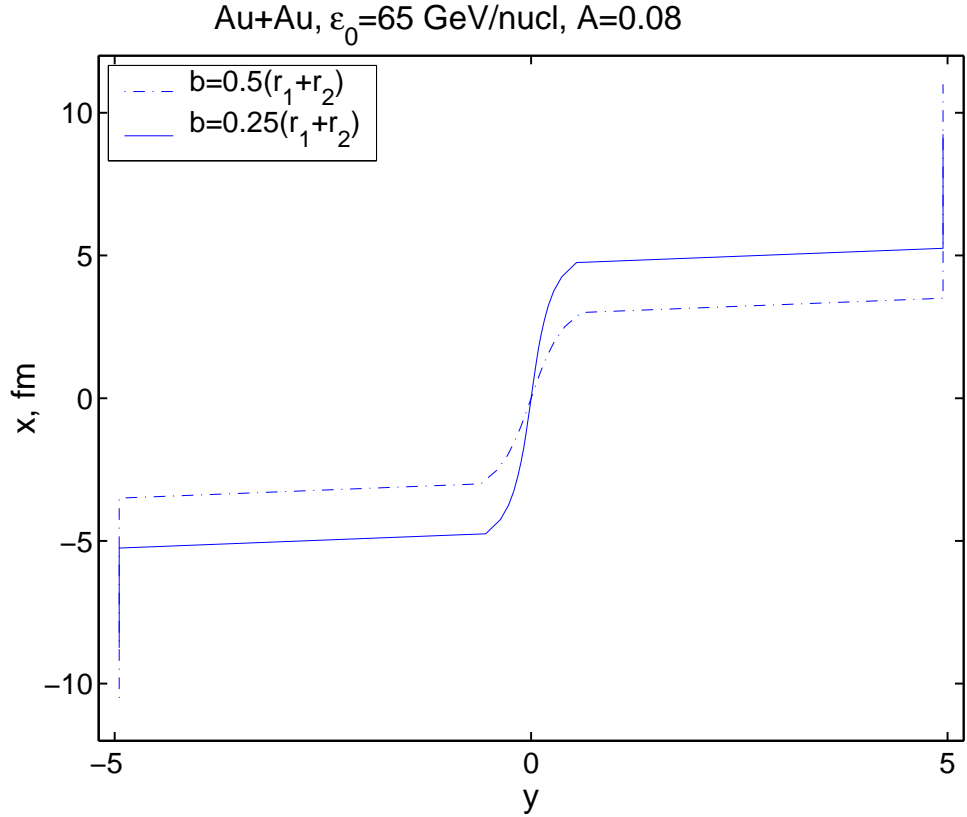}
\vspace{0.6cm}
\caption[]{The rapidity profile of streaks in the reaction plane 
for Au+Au collision at 
$\varepsilon_0=65\ GeV/nucl$,
$A=0.065$, $y=0$. The rapidities of the final streaks in CRF
are calculated according to eq. (\ref{eq66}). Our 
profiles are in agreement with schematic 
sketch in paper \cite{SSVWX}.}

\label{rap_dist}
\end{figure}

\section{$\!\!\!\!\!\!$: The analytical solution of the model}
\label{app2}
For $x^\pm>x_0$ we should solve eqs. (\ref{eq26}), based on 
boundary conditions (\ref{eq45}).
Eqs. (\ref{eq26}) leads to the system of equations:
\beq
\partial_-\left(h_{1+}e^{-2y_1}\right)+\alpha\partial_+h_{2+}=
-2\sigma n_0 e^{y_0} + 2\sigma n_0 e^{y_0}e^{-2y_2}\ ,
\eeq{eq46}
\beq
\alpha\partial_-h_{1+}+\partial_+\left(h_{2+}e^{2y_2}\right)=
2\sigma n_0 e^{y_0}e^{2y_1} - 2\sigma n_0 e^{y_0}\ ,
\eeq{eq47}
where $\alpha=(1-c_0^2)/(1+c_0^2)$.
It is clear, that in both equations there are two terms depending 
on independent variables, so the solution will contain two undefined 
constants.  The next step is to take eqs. (\ref{eq46}, \ref{eq47}) 
at the values $x^+=x_0$ and $x^-=x_0$. 
\beq
h_{1+}=e^{2y_1}\left(
e_1(t_0)(1+c_0^2)e^{2y_0}-a_2(x^--x_0)\right) \ ,
\eeq{eq48}
\beq
\alpha \partial_+h_{2+}=c_2+2\sigma n_0e^{y_0}\left(
e^{-2y_2}-e^{-2y_0}\right) \ ,
\eeq{eq49}
\beq
h_{2+}=e^{-2y_2}\left(
e_2(t_0)(1+c_0^2)e^{2y_0}-a_1(x^+-x_0)\right) \ ,
\eeq{eq50}
\beq
\alpha \partial_-h_{1+}=c_1+2\sigma n_0e^{y_0}\left(
e^{2y_1}-e^{-2y_0}\right) \ ,
\eeq{eq51}
where we introduced new notations
\beq
a_1=c_1+4 \sigma n_0 \sinh {y_0}
\ , \quad  
a_2=c_2+4\sigma n_0 \sinh {y_0} \ 
\eeq{eq52}
and two new constants 
\beq
c_1=\alpha (h_{1+})'|_{x_0}\ , \quad  c_2=\alpha (h_{2+})'|_{x_0} \ ,
\eeq{eq53}
which will be estimated by assuming a linear development for the enthalpy
densities, $h_{1+}$ and  $h_{2+}$, from $t=0\ (x_\pm=|z(0)|)$ to
$t=t_0\ (x_\pm=x_0)$.
\beq
c_i=\alpha ((1+c_0^2)e_i(t_0)-e_0)/2t_0 \ .
\eeq{eq61}

The complete analytical solution found to be 
\beq
e^{(-)^{i+1}2y_i}=-\frac{d_i}{b_i}+
\left(\frac{d_i}{b_i}+e^{-2y_0}\right)
\left(1-\frac{x^i-x_0}{\tau_i}\right)^{-\frac{b_i}{\alpha a_j}} \ ,
\eeq{eq54a}
\beq
h_{i+}=e^{(-)^{i+1}2y_i}e_i(t_0)(1+c_0^2)e^{2y_0}
\left(1-\frac{x^i-x_0}{\tau_i}\right)\ ,
\eeq{eq55a}
\beq
 n_i= n_0e^{y_0}e^{(-)^{i+1}y_i} \ ,
\eeq{eq56a}
where $x^1=x^-,\ x^2=x^+$, $i,j=1,2\ ,\ i\not=j$,
\beq
b_i=\alpha a_j+2\sigma n_0e^{y_0} \ ,
\eeq{eq57.b}
\beq
d_i=c_i-2\sigma n_0e^{-y_0} \ ,
\eeq{eq57.d}
\beq
\tau_i=\frac{e_i(t_0)(1+c_0^2)}{e^{-2y_0}a_j} \ .
\eeq{eq58}

\section{$\!\!\!\!\!\!$: Transformation to the lightcone coordinates}
\label{app3}
In this Appendix we present the transformation matrixes between $(t,z)$
and $(+,-)$ coordinates. Indexes $i,j,k$ run for $0,3$, indexes $\alpha,
\beta,\gamma$ are used for $+,-$.

The transformation of the coordinate system is
\beq
x^\pm=t\pm z\ ,
\eeq{t1}
thus, for all contravariant vectors we have the same: 
\beq
V^\pm=V^0\pm V^3\ .
\eeq{t1a}
For $(t,z)$ coordinates we have
\beq
g_{ik}=g^{ik}=\left(
\begin{array}{cc}
1& 0\\
0& -1
\end{array}\right)\ .
\eeq{t2}
Then
\beq
g_{\alpha\beta}=g^{\alpha\beta}=\left(
\begin{array}{cc}
0& 1\\
1& 0
\end{array}\right)\ ,
\eeq{t3}
\beq
g^i_{.\alpha}=g_i^{.\alpha}={1\over 2}\left(
\begin{array}{cc}
1& 1\\
1& -1
\end{array}\right)\ ,
\eeq{t4}
\beq
g^\alpha_{.i}=g_\alpha^{.i}=\left(
\begin{array}{cc}
1& 1\\
1& -1
\end{array}\right)\ .
\eeq{t4b}

\beq
T^{\alpha\beta}=g^\alpha_{.i} T^{ij} g_j^{.\beta}\ ,
\eeq{t5}
so,
\beq
T^{++}={1\over 2}\left(T^{00}+T^{03}+T^{30}+T^{33}\right)\ ,
\eeq{t5a}
\beq
T^{+-}={1\over 2}\left(T^{00}-T^{03}+T^{30}-T^{33}\right)\ ,
\eeq{t5b}
\beq
T^{-+}={1\over 2}\left(T^{00}+T^{03}-T^{30}-T^{33}\right)\ ,
\eeq{t5c}
\beq
T^{--}={1\over 2}\left(T^{00}-T^{03}-T^{30}+T^{33}\right)\ .
\eeq{t5d}

The backward transformation is
\beq
T^{ij}=g^i_{.\alpha} T^{\alpha\beta} g^\beta_{.j} ,
\eeq{t6}
so,
\beq
T^{00}={1\over 2}\left(T^{++}+T^{+-}+T^{-+}+T^{--}\right)\ ,
\eeq{t6a}
\beq
T^{03}={1\over 2}\left(T^{++}-T^{+-}+T^{-+}-T^{--}\right)\ ,
\eeq{t6b}
\beq
T^{30}={1\over 2}\left(T^{++}+T^{+-}-T^{-+}-T^{--}\right)\ ,
\eeq{t6c}
\beq
T^{33}={1\over 2}\left(T^{++}-T^{+-}-T^{-+}+T^{--}\right)\ .
\eeq{t6d}

\vfill \eject


\begin{thebibliography}{99}
\bibitem{FO1}
      L.P. Csernai,  Zs. L\'az\'ar and D. Moln\'ar,
      {\it Heavy Ion Phys.} {\bf 5} (1997) 467.

\bibitem{FO2}
      Cs. Anderlik, Z.I. L\'az\'ar, V.K. Magas, L.P. Csernai, H. St\"ocker
      and W. Greiner, {\it Phys. Rev.} {\bf C 59} (1999)  388.
      (nucl-th/9808024)

\bibitem{FO2a}
  Cs. Anderlik, L.P. Csernai, F. Grassi, W. Greiner, Y. Hama, T. 
Kodama, Zs. Lazar, V. Magas and H. Stoecker,    
{\it  Phys. Rev.} {\bf C 59} (1999) 3309. (nucl-th/9806004)

\bibitem{FO3}
 V.K. Magas, Cs. Anderlik, L.P. Csernai, F. Grassi,
 W. Greiner, Y. Hama, T. Kodama, Zs. L\' az\' ar and 
 H. St\" oker, 
{\it \ Heavy Ion Phys.} {\bf 9} (1999) 193. (nucl-th/9903045)

\bibitem{FO3a}
  Cs. Anderlik, L.P. Csernai, F. Grassi, W. Greiner, Y. Hama, T. 
Kodama, Zs. Lazar, V. Magas and H. Stoecker,    
{\it  Phys. Lett.} {\bf B 459} (1999) 33. (nucl-th/9905054)

\bibitem{FO3b}
 V.K. Magas, Cs. Anderlik, L.P. Csernai, F. Grassi,
 W. Greiner, Y. Hama, T. Kodama, Zs. L\' az\' ar and 
 H. St\" oker, 
{\it \ Nucl. Phys.} {\bf A 661} (1999) 596. (nucl-th/0001049)

\bibitem{RG96}
D.H. Rischke, M. Gyulassy, {\it \ Nucl. Phys.} {\bf A608} (1996) 479.

\bibitem{LJP_qm}
F. Laue, talk at the Quark Matter 2001, Stony Brook, Janyary 15-20, 2001; 
S. Johnson, talk at the Quark Matter 2001, Stony Brook, Janyary 15-20, 2001; 
S. Panitkin, talk at the Quark Matter 2001, Stony Brook, Janyary 15-20, 2001. 

\bibitem{CM95}
L.P. Csernai, I.N. Mishustin, {\it \ Phys. Rev. Lett.} {\bf 74} (1995)
5005. 

\bibitem{A78}
 A.A. Amsden, A.S. Goldhaber, F.H. Harlow and J.R. Nix,
  {\it \ Phys. Rev. } {\bf C17} (1978) 2080.

\bibitem{C82}
L.P. Csernai, I. Lovas, J. Maruhn, A. Risenhauer, J. Zimanyi,
W. Greiner,  {\it \ Phys. Rev. } {\bf C26} (1982) 149.

\bibitem{bsd00}
J. Brachmann, S. Soff, A. Dumitru, H. St\" oker, J.A. Maruhn, W. Greiner, 
D.H. Rischke,
L. Bravina, {\it \ Phys. Rev. } {\bf C61} (2000) 024909.

\bibitem{bnk84}
T.S. Bir\'o, H.B. Nielsen, J. Knoll,  {\it \ Nucl. Phys.} {\bf B245} (1984) 
449.

\bibitem{S95}
H. Sorge, {\it \ Phys. Rev. } {\bf C52} (1995) 3291.

\bibitem{WA96}
K. Werner, J. Aichelin, {\it \ Phys. Rev. Lett.} {\bf 76} (1996) 1027.

\bibitem{ABP93}
N.S. Amelin, M.A. Braun, C. Pajares,
{\it \ Phys. Lett.} {\bf B306} (1993) 312, {\it \ Z. Phys.} {\bf C63} 
(1994) 507.

\bibitem{ASC91pl} 
    N. S. Amelin, E.F. Staubo, L.P. Csernai, V.D. Toneev, K.K. Gudima and
    D.D. Strottman,
    {\it Phys. Lett.} {\bf B261}  (1991) 352.

\bibitem{ASC91prl} 
    N. S. Amelin, E.F. Staubo, L.P. Csernai, V.D. Toneev, K.K. Gudima and
    D.D. Strottman,
    {\it Phys. Rev. Lett.} {\bf 67} (1991) 1523.

\bibitem{ACS92}
    N.S. Amelin, L.P. Csernai, E.F. Staubo, and D. Strottman
    {\it Nucl. Phys. } {\bf A544} (1992) 463c.

\bibitem{GC86}
M. Gyulassy, L.P. Csernai, {\it \ Nucl. Phys.} {\bf A460} (1986) 723.

\bibitem{EK99}
K.J. Eskola, K. Kajantie, P.V. Ruuskanen, K. Tuominen, 
{\it \ Nucl. Phys.} {\bf B570} (2000) 379. 

\bibitem{DG00}
A. Dumitru, M. Gyulassy,
{\it \ Phys. Lett. } {\bf B 494} (2000) 215. 

\bibitem{QM01}
P. Steinberg, talk at the Quark Matter 2001, Stony Brook, Janyary 15-20, 2001; 
R. Snellings, talk at the Quark Matter 2001, Stony Brook, Janyary 15-20, 2001; 
A. Drees, talk at the Quark Matter 2001, Stony Brook, Janyary 15-20, 2001. 


\bibitem{MCS00} 
V.K. Magas, L.P. Csernai, D.D. Strottman, Proceedings of New 
Trends in High-Energy Physics, Yalta (Crimea), Ukraine,
May 27 - June 4, 2000, p. 93. (nucl-th/0009049) 

\bibitem{CAM00}
L.P. Csernai, Cs. Anderlik, V.K. Magas,
presented at the 
Symposium on Fundamental in Elementary Matter, Bad Honnef, Germany, 
September 25-29, 2000. (nucl-th/0010023) 
 
\bibitem{MCS00-1}
V. Magas, L.P. Csernai, D. Strottman,
presented at the ISMD 2000 - XXXth International Symposium on
Multiparticle Dynamics, Tihany, Lake Balaton, Hungary, 
October 9-15, 2000. (hep-ph/0101125)

\bibitem{KM85} K. Kajantie, T. Matsui,    
{\it Phys. Lett.} {\bf B164}  (1985) 373.
 

\bibitem{CR}
L.P. Csernai, D. R\"ohrich,
{\it \ Phys. Lett.} {\bf B458} (1999) 454. (nucl-th/9908034)

\bibitem{P_qm} P. Huovinen, talk at the 
Quark Matter 2001, Stony Brook, Janyary 15-20, 2001. 

\bibitem{SSVWX}
R.J.M. Snellings, H. Sorge, S.A. Voloshin, F.Q. Wang, N. Xu, 
{\it Phys. Rev. Lett.} {\bf 84} (2000) 2803-2805. (nucl-ex/9908001)

\end{thebibliography}
\end{document}